\documentclass[letter,twocolumn]{jpsj3}
\usepackage{txfonts}
\usepackage{color}

\title{
Third boundary 
of the Shastry-Sutherland Model by Numerical Diagonalization
}
\catcode`\@=11
\def\simle{\mathrel{\mathpalette\@versim<}}   
\def\simge{\mathrel{\mathpalette\@versim>}}   
\def\@versim#1#2{\lower2.5pt\vbox{\baselineskip0pt \lineskip-.5pt
   \ialign{$\m@th#1\hfil##\hfil$\crcr#2\crcr\sim\crcr}}}
\catcode`\@=12

\author{Hiroki Nakano$^1$
 and T\^oru Sakai$^{1,2}$}
\inst{$^1$
Graduate School of Material Science, 
University of Hyogo,
Kamigori, 
Hyogo 678-1297, Japan \\
$^2$
National Institutes for Quantum and Radiological Science and Technology, 
SPring-8
Sayo, Hyogo 679-5148, Japan 
}

\abst{
The Shastry-Sutherland model
\--- 
the $S=1/2$ Heisenberg antiferromagnet 
on the square lattice accompanied by orthogonal dimerized interactions 
\---  
is studied 
by the numerical-diagonalization method. 
Large-scale calculations provide results for larger clusters 
that have not been reported yet. 
The present study successfully captures the phase boundary between 
the dimer and plaquette-singlet phases  
and 
clarifies that the spin gap increases once 
when the interaction forming the square lattice is increased 
from the boundary.  
Our calculations strongly suggest that 
in addition to the edge of the dimer phase given by 
$J_{2}/J_{1}\sim 0.675$ and 
the edge of the N$\acute{\rm e}$el-ordered phase given by
$J_{2}/J_{1}\sim 0.76$, 
there exists a third boundary ratio $J_{2}/J_{1}\sim 0.70$ 
that divides the intermediate region into two parts, 
where $J_{1}$ and $J_{2}$ denote dimer and
square-lattice interactions, respectively.
}

\begin{document}
\maketitle


It is well known that frustration in magnetic materials 
enables exotic quantum states to be realized. 
However, not many such quantum states are obtained 
in a mathematically rigorous form.  
The Shastry-Sutherland model\cite{Shastry_Sutherland_Physica1981} 
is a member of the family of mathematical models 
in which not all but only some of the eigenstates are exactly obtained 
\cite{Majumdar_Ghosh_1969,Majumdar_1970,Caspers_1982,
AKLT_PRL1987,Lange_ZPhys1994,Long_Siak_1993,HN_VBS_PRB1996,
HN_JPSJ1997,HN_JPSJ1998,Tonooka_JPSJ2007}. 
Among such systems, the Shastry-Sutherland model 
became important after a good candidate material, SrCu$_{2}$(BO$_{3}$)$_{2}$, 
was discovered\cite{Kageyama_PRL1999,SMiyahara_KUeda_PRL1999}. 
The discovery was followed 
by extensive theoretical and experimental studies. 

Between the region with the exact dimer ground state
and
the weakly frustrated region
with the typical N$\acute{\rm e}$el-ordered ground state, 
the existence of the plaquette-singlet phase 
was pointed out in Ref.~\ref{AKoga_NKawakami_PRL2000}. 
Various approaches\cite{YFukumoto_JPSJ2000,
Lauchli_PRB2002,Lou_arXiv1212_1999,
Corboz_Mila_PRB2013,Wang_Batista_PRL2018}
have theoretically attempted to clarify
the behavior of the system in the intermediate region. 
In the numerical-diagonalization studies, unfortunately, 
the maximum of the treated system sizes 
\--- 
32 spin sites 
in Ref.~\ref{Lauchli_PRB2002} to the best of our knowledge 
\--- 
was not so large. 
On the other hand, 
the pressure dependence of the spin gap was observed 
experimentally\cite{HOhta_JPhysChem2015,Zayed_NatPhys2017,
TSakurai_JPSJ87_ESR_HP}. 
Among these studies, Ref.~\ref{TSakurai_JPSJ87_ESR_HP}, 
employing an electron spin resonance study under 
high pressure and high field, recently reported 
the behavior of the spin gap 
around the phase transition at the edge of the dimer phase. 
This experimental result becomes a significant motivation 
to theoretically clarify the behavior of the spin gap 
around the intermediate region 
from the numerical-diagonalization 
calculations for even larger systems. 

Under the circumstances, the purpose of the present paper is 
to report numerical-diagonalization results of even larger systems 
and 
to examine the behavior of the system 
around the intermediate region 
between the dimer and N$\acute{\rm e}$el-ordered regions. 
The present study provides results 
for 36-site and 40-site systems in addition to smaller systems 
to deepen our understanding of this system. 
In particular, numerical diagonalizations 
of the 40-site system require large-scale parallel calculations 
in an appropriate supercomputer. 
We successfully detect the edge of the dimer phase 
and the edge of the N$\acute{\rm e}$el-ordered phase 
from the results of the two large sizes.  
Between the two edges, we additionally detect 
the third ratio of the boundary dividing the intermediate region 
into two parts, each state of which has characteristics 
of a correlation function that are different from each other.


\begin{figure}[tb]
\begin{center}
\includegraphics[width=8cm]{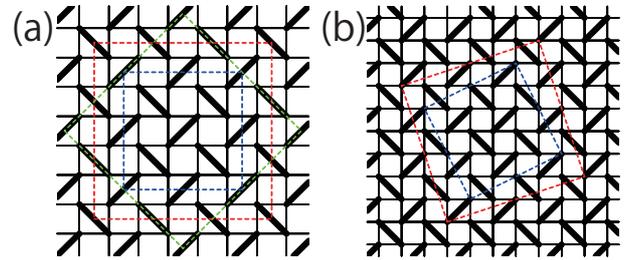}
\end{center}
\caption{(Color) 
Lattice structure of the orthogonal dimer system, 
namely, the Shastry-Sutherland model. 
Thick solid lines and thin solid lines denote 
bonds for $J_1$ and $J_2$, respectively. 
Panel (a) shows the finite-size clusters 
of $N_{\rm s}=16$, 32, and 36 by blue, green, and red dotted lines, 
respectively.  
Panel (b) shows the finite-size clusters 
of $N_{\rm s}=20$ and 40 by blue and red dotted lines, 
respectively.  
}
\label{fig1}
\end{figure}

The Hamiltonian studied here is given by 
\begin{equation}
{\cal H}
=
\sum_{\langle i ,j\rangle : \ {\rm orthogonal~dimer}} J_{1} 
\mbox{\boldmath $S$}_{i}\cdot\mbox{\boldmath $S$}_{j} 
+
\sum_{\langle i ,j\rangle : \ {\rm square~lattice}} J_{2} 
\mbox{\boldmath $S$}_{i}\cdot\mbox{\boldmath $S$}_{j} 
. 
\label{Hamiltonian}
\end{equation}
Here, $\mbox{\boldmath $S$}_{i}$ 
represents the $S=1/2$ spin operator at site $i$. 
We consider the case of an isotropic interaction 
in spin space in this study. 
Site $i$ is assumed to characterize the vertex 
of the square lattice. 
The number of spin sites is denoted by $N_{\rm s}$. 
The first term of Eq.~(\ref{Hamiltonian}) denotes 
orthogonal dimer interactions represented 
by thick solid bonds in Fig.~\ref{fig1}. 
The second term of Eq.~(\ref{Hamiltonian}) represents 
interactions forming the square lattice
represented by thin solid bonds in Fig.~\ref{fig1}. 
We consider that the two interactions between the two spins 
are antiferromagnetic, 
namely, $J_{1} > 0$ and $J_{2} > 0$. 
Energies are measured in units of $J_{1}$; 
hereafter, we set $J_{1}=1$. 
We denote the ratio $J_{2}/J_{1}$ by $r$.  
Note here that when $r=0$, the system is an assembly of 
isolated dimerized-spin models,  
whereas the system is reduced 
to the $S=1/2$ Heisenberg antiferromagnet 
on the ordinary square lattice 
in the limit $r\rightarrow\infty$. 

We treat finite-size clusters with system size $N_{\rm s}$ 
under the periodic boundary condition. 
In this study, 
$N_{\rm s}=16$, 20, 32, 36, and 40 are treated;  
finite-size clusters are shown in Fig.~\ref{fig1}. 
Note here that $N_{\rm s}/4$ is an integer 
and that all the clusters are regular squares, 
although the squares for $N_{\rm s}=20$, 32, and 40 are tilted. 
The regular-square clusters help us 
capture well the two dimensionality of the present system. 

We carry out our numerical diagonalizations 
on the basis of the Lanczos algorithm 
to obtain the lowest energy of ${\cal H}$ 
in the subspace belonging to $\sum _j S_j^z=M$. 
Note here that the $z$-axis 
is taken as the quantized axis of each spin.  
It is widely believed that 
numerical-diagonalization calculations are unbiased. 
Thus, one can obtain reliable information about the system. 
The energy is denoted by $E(N_{\rm s},M)$, 
where $M$ is an integer;  
in particular, we calculate the cases $M=0$ and $M=1$ 
because our attention is focused primarily 
on the behavior of the spin gap given by
\begin{equation}
\Delta=E(N_{\rm s},1) - E(N_{\rm s},0) .  
\label{gap_determining_finite_N}
\end{equation}
Some of the Lanczos diagonalizations were carried out 
using MPI-parallelized code that was originally 
developed in the study of Haldane gaps\cite{HNakano_HaldaneGap_JPSJ2009}. 
The usefulness of our program was confirmed in large-scale 
parallelized calculations\cite{HNakano_kgm_gap_JPSJ2011,
HNakano_s1tri_LRO_JPSJ2013,HN_TSakai_kgm_1_3_JPSJ2014,
HN_TSakai_kgm_S_JPSJ2015,HN_YHasegawa_TSakai_dist_s
huriken_JPSJ2015,HN_TSakai_dist_tri_JPSJ2017,
HN_TSakai_tri_NN_JPSJ2017,HN_TSakai_kgm45_JPSJ2018,
YHasegawa_HN_TSakai_dist_shuriken_PRB2018,
TSakai_HN_ICM018,HN_TSakai_S2HaldaneGap_JPSJ2018}. 
Note here that the largest-scale
calculations in this study 
have been carried out using either the K computer or Oakforest-PACS. 


\begin{figure}[tb]
\begin{center}
\includegraphics[width=8cm]{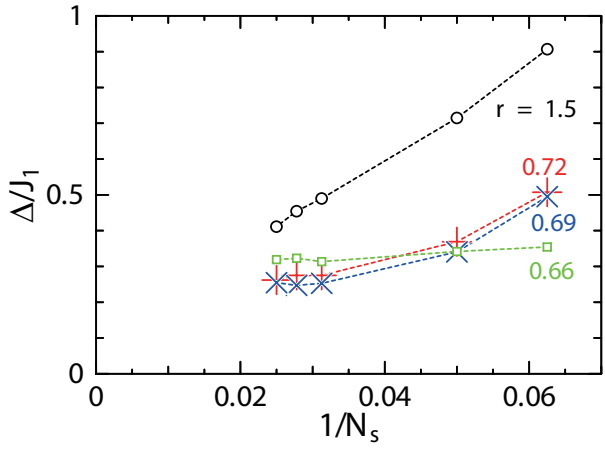}
\end{center}
\caption{(Color) 
Finite-size energy differences $\Delta$ 
of the cases $N_{\rm s}=16$, 20, 32, 36, and 40 
for some representative $r$ as a function of $1/N_{\rm s}$.  
Black circles, red pluses, blue crosses, and green squares 
denote results for $r=1.5$, 0.72, 0.69, and 0.66, respectively.
}
\label{fig2}
\end{figure}

Now, let us observe the $N_{\rm s}$-dependence of $\Delta/J_{\rm 1}$ 
for some representative cases of $r$; 
the results are depicted in Fig.~\ref{fig2}. 
One finds that for $r=1.5$, $\Delta/J_{\rm 1}$ significantly decreases 
as $N_{\rm s}$ is increased. 
The decreasing behavior of $\Delta/J_{\rm 1}$ is consistent with 
that in the gapless N$\acute{\rm e}$el-ordered phase. 
Our results for $r=1.5$ suggest 
an almost linear dependence on $1/N_{\rm s}$. 
On the other hand, for $r=0.72$ and 0.69,  
$\Delta/J_{\rm 1}$ decreases with increasing $N_{\rm s}$ for small $N_{\rm s}$ 
but shows only a very weak $N_{\rm s}$-dependence for large $N_{\rm s}$. 
For $r=0.66$, 
$\Delta/J_{\rm 1}$ finally becomes almost constant for all ranges 
of $N_{\rm s}$. 
From these observations, it is considered 
that clusters with $N_{\rm s}=36$ and 40 capture well 
the behavior of large systems approaching the thermodynamic limit. 
Therefore, focusing our attention on the results of $N_{\rm s}=36$ and 40, 
we hereafter investigate the behavior of the present system. 

\begin{figure}[tb]
\begin{center}
\includegraphics[width=8cm]{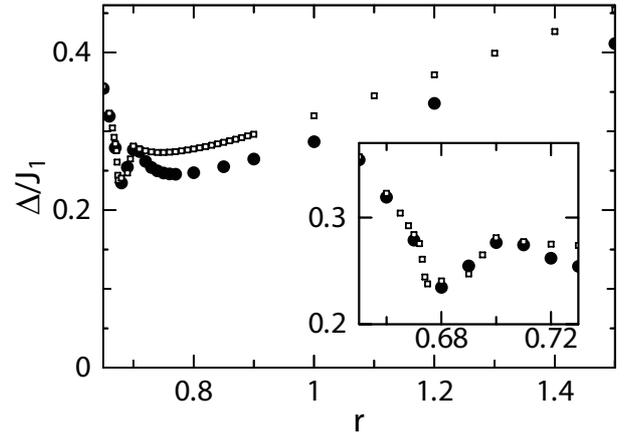}
\end{center}
\caption{
$r$-dependence of the spin gap 
for $N_{\rm s} = 36$ and 40. 
Open squares and closed circles denote results 
for $N_{\rm s} = 36$ and 40, respectively. 
Inset is a zoom-in view of the region of $r$ 
around $r\sim 0.7$. 
}
\label{fig3}
\end{figure}
Next, let us observe the $r$-dependence 
of the spin gap $\Delta$ for finite-size clusters in detail; 
the results of $N_{\rm s}=36$ and 40 are depicted in Fig.~\ref{fig3}. 
First, one finds, in the region up to $r\sim 0.67$, 
that $\Delta/J_{1}$ gradually decreases as $r$ is increased 
and 
that data for the two sizes agree well with each other. 
The agreement strongly suggests that 
the system-size dependence has already become weak 
and 
that finite-size results almost agree 
with the corresponding values for the thermodynamic limit. 
Next, in the region from $r\sim 0.68$ to $r\sim 0.70$, 
on the other hand, $\Delta/J_{1}$ gradually increases 
with increasing $r$. 
The good agreement of data for the two sizes is still maintained, 
although the $r$-dependence of whether it increases or decreases 
has been changed. 
In the region above $r\sim 0.70$, 
$\Delta/J_{1}$ decreases once but increases again 
as $r$ is increased. 
The upturn of $\Delta/J_{1}$ is observed 
for both $N_{\rm s}=36$ and 40. 
The significant characteristic in this region 
is that there appears a considerable system-size dependence: 
that is, $\Delta/J_{1}$ for $N_{\rm s}=40$ is smaller than
that for $N_{\rm s}=36$ at a given $r$. 
To find whether the nonzero spin gap exists or 
is absent in the thermodynamic limit, 
we will need to carry out further analysis. 

\begin{figure}[tb]
\begin{center}
\includegraphics[width=8cm]{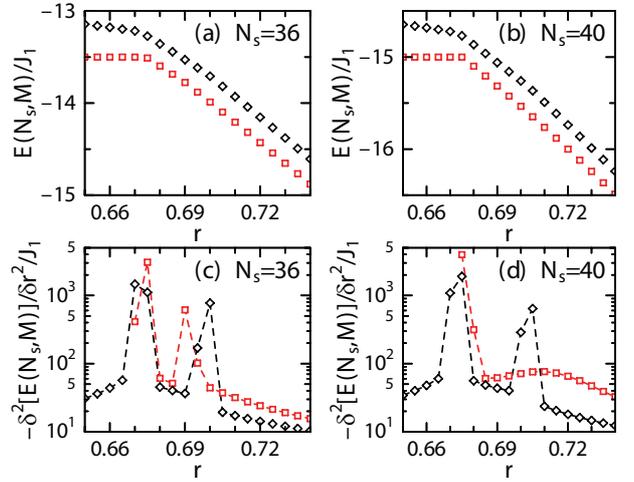}
\end{center}
\caption{(Color) 
Energy-level scheme and second derivatives of the energies 
with respect to the ratio $r$. 
Panels (a) and (b) show results of the energy-level scheme 
for $N_{\rm s} = 36$ and 40, respectively. 
Panels (c) and (d) show results of the second derivatives 
for $N_{\rm s} = 36$ and 40, respectively. 
Squares and diamonds denote results for $M=0$ and $M=1$, respectively.
}
\label{fig4}
\end{figure}

To deepen our understanding of the behavior of $\Delta$ 
showing a complex dependence of decreases and increases,  
next, let us observe the $r$-dependence of $E(N_{\rm s},M)$; 
the results are depicted in Fig.~\ref{fig4}(a) and (b). 
First, one easily finds that 
$E(N_{\rm s},0)$ in the region up to $r\sim 0.67$ is constant.
The constant value corresponds to the eigenenergy 
of the rigorous dimer ground state.
This clearly indicates that 
this region corresponds to the dimer phase with the exact ground state. 
One finds, from the results of $E(N_{\rm s},0)$ above $r \sim 0.675$, that 
another state different from the dimer state becomes the ground state. 
No significant size dependence is observed 
with respect to the boundary ratio $r \sim 0.675$. 
This result agrees well with previously known estimates 
of the phase boundary: 
$r=0.677$ in Ref.~\ref{AKoga_NKawakami_PRL2000},
$r=0.678$ in Ref.~\ref{Lauchli_PRB2002},
$r=0.687$ in Ref.~\ref{Lou_arXiv1212_1999}, and 
$r=0.675$ in Ref.~\ref{Corboz_Mila_PRB2013}. 
In order to capture the behavior in the region $r \simge 0.675$, 
we evaluate a numerical second derivative given by 
$-\delta^{2}[E(N_{\rm s},M)]/\delta r^{2}=[2E(N_{\rm s},M)|_{r}
-E(N_{\rm s},M)|_{r+\delta r}-E(N_{\rm s},M)|_{r -\delta r}]/(\delta r)^{2}$; 
the results are depicted in Fig.~\ref{fig4}(c) and (d). 
It is known that the analysis based on second derivatives
is useful to detect the boundaries 
of a target system\cite{You_second_deri_PRE2007,
Albuquerque_second_deri_PRB2010,HN_TSakai_tri_NN_JPSJ2017}. 
In Fig.~\ref{fig4}(c) and (d), 
the second derivatives can appropriately capture 
the discontinuity at $r \sim 0.675$ for both $M=0$ and $M=1$. 
In addition, the second derivatives 
for $N_{\rm s}=36$ 
show another discontinuity around $r\sim 0.69$-0.70. 
For $N_{\rm s}=40$, the second derivative of $M=1$ also shows 
a discontinuity 
around $r\sim 0.70$; 
that of $M=0$ does not show a discontinuity but 
it shows a peak at $r\sim 0.71$ instead. 
The behavior around $r\sim 0.70$ is consistent with 
the result $r=0.702$ for the edge of the plaquette-singlet phase 
reported in the $N_{\rm s}=32$ diagonalization study 
in Ref.~\ref{Lauchli_PRB2002}. 
On the other hand, the present result of $r\sim 0.70$ 
differs from 
$r=0.86$ in Ref.~\ref{AKoga_NKawakami_PRL2000},
$r=0.75$ in Ref.~\ref{Lou_arXiv1212_1999}, and 
$r=0.765(15)$ in Ref.~\ref{Corboz_Mila_PRB2013}
as results for the edge of the plaquette-singlet phase. 
The difference will be examined later. 
From Fig.~\ref{fig4}, therefore, one can understand that
the observed changes in the dependence of $\Delta$ 
are due to the energy-level structure 
of both $E(N_{\rm s},0)$ and $E(N_{\rm s},1)$.  

\begin{figure}[tb]
\begin{center}
\includegraphics[width=8cm]{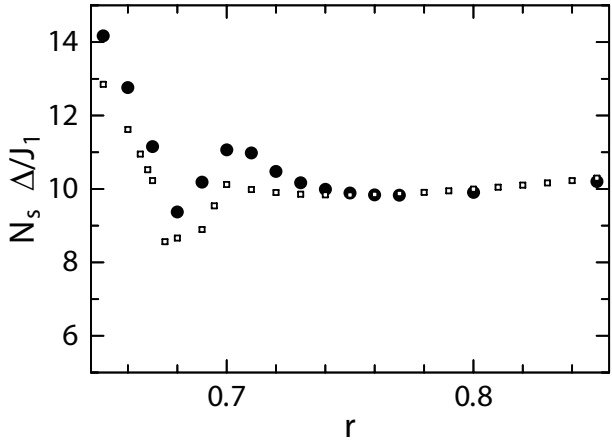}
\end{center}
\caption{
$r$-dependence of the product of the system size and the spin gap 
for $N_{\rm s} = 36$ and 40. 
Open squares and closed circles denote results 
for $N_{\rm s} = 36$ and 40, respectively. 
}
\label{fig5}
\end{figure}
Next, let us examine the system-size dependence of the spin gap
in the region of large $r$. 
When $r$ is infinitely large, the system is reduced to 
the simple square-lattice antiferromagnet, showing that 
the spin excitation is gapless 
owing to the existence of the N$\acute{\rm e}$el order.  
As a means of distinguishing  
whether the system is gapped or gapless, 
the method of observing the product of the system size and the spin gap 
is known. 
This method was successfully used in the study of 
the plateau \--- the gap under the magnetic field \--- 
at the one-third height of the saturation 
in the triangular-lattice Heisenberg antiferromagnet 
with next-nearest-neighbor interactions\cite{HN_TSakai_tri_NN_JPSJ2017}. 
The results of this analysis for the present system 
with $N_{\rm s}=36$ and 40 are depicted in Fig.~\ref{fig5}. 
One clearly finds that the results from the two sizes 
agree with each other in the region down to $r\sim 0.75$. 
When $r$ is further decreased, the results of $N_{\rm }=40$ 
clearly become larger than those of $N_{\rm }=36$. 
The agreement in the behavior of $N_{\rm s} \Delta/J_{1}$ 
in the region of large $r$ suggests that  the finite-size spin gap 
in this region exhibits $\Delta \propto 1/N_{\rm s}$, 
which means that the system is gapless. 
In the region below $r\sim 0.75$, on the other hand, 
the finite-size spin gap does not have 
the dependence $\Delta \propto 1/N_{\rm s}$.  
Although the $N_{\rm s}$-dependence of $\Delta$ 
in the region between $r\sim 0.71$ and $r\sim 0.75$ is unclear 
at the present stage, there are two possible scenarios. 
One is that the system is gapped without any long-range orders. 
The other is that the system is gapless, but 
the $N_{\rm s}$-dependence of $\Delta$ is different 
from $\Delta \propto 1/N_{\rm s}$ 
corresponding to the N$\acute{\rm e}$el-ordered phase.

\begin{figure}[tb]
\begin{center}
\includegraphics[width=8cm]{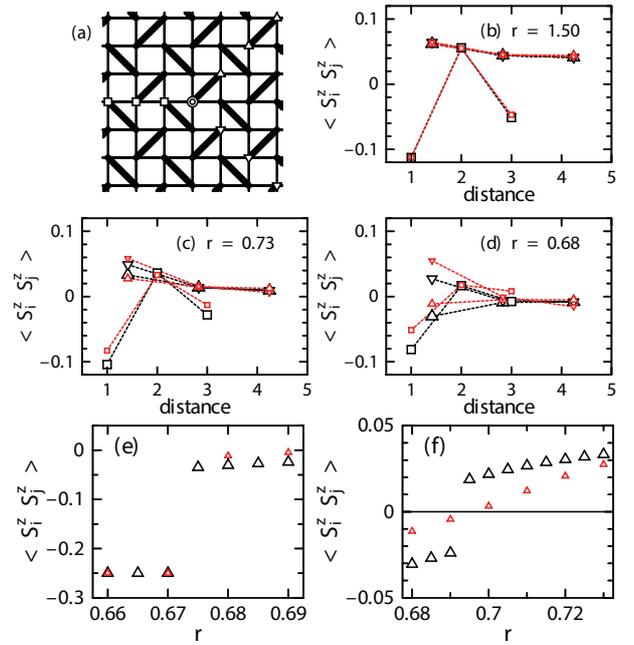}
\end{center}
\caption{(Color) 
Correlation functions $\langle S_{i}^{z} S_{j}^{z} \rangle$. 
Panel (a) shows the positions of site $i$ and $j$. 
For a given site $i$ denoted by the double circle, 
$j$ is taken in three directions shown by squares, 
triangles, and inversed triangles. 
Panels (b), (c), and (d) show results 
for $\langle S_{i}^{z} S_{j}^{z} \rangle$ 
as a function of the distance between $i$ and $j$ 
for $r=1.50$, 0.73, and 0.68, respectively. 
Panels (e) and (f) show the $r$-dependence 
of $\langle S_{i}^{z} S_{j}^{z} \rangle$ 
for the shortest-distance pair 
along the direction represented by triangles.
Black and red symbols denote results 
for $N_{\rm s} = 36$ and 40, respectively. 
}
\label{fig6}
\end{figure}
To capture the change from the N$\acute{\rm e}$el-ordered phase 
to the plaquette-singlet phase, 
let us observe correlation functions in the ground state, 
namely, $\langle S_{i}^{z} S_{j}^{z} \rangle$; 
the results for both $N_{\rm s}=36$ and $N_{\rm s}=40$ 
are depicted in Fig.~\ref{fig6}. 
The case $r=1.50$ in Fig.~\ref{fig6}(b) is a typical one 
for the N$\acute{\rm e}$el-ordered phase. 
All the results shown by triangles and inversed triangles 
are positive; this feature is explained by the fact that 
both $i$ and $j$ for a measured pair are in a common sublattice 
among the two sublattices of the N$\acute{\rm e}$el-ordered state. 
The results shown by the triangles and the inversed triangles 
also indicate a gradual decay as the distance is increased. 
The results shown by the squares indicate alternating signs. 
This behavior suggests the staggered nature 
of the N$\acute{\rm e}$el-ordered state. 
Therefore, the characteristics of the N$\acute{\rm e}$el-ordered state 
are well captured in Fig.~\ref{fig6}(b).  
The results in Fig.~\ref{fig6}(d), on the other hand, 
are completely different from those in Fig.~\ref{fig6}(b). 
Among the results shown by the triangles and the inversed triangles, 
only the shortest-distant datum by the inversed triangle is positive, 
and the rest are negative. 
Absolute values of $\langle S_{i}^{z} S_{j}^{z} \rangle$ 
for distances larger than two are very small. 
These behaviors of the correlation functions 
are different from those of the N$\acute{\rm e}$el-ordered state, 
but are consistent with those of the plaquette-singlet state.    
In this state, each plaquette singlet is located at a local square
involving the $J_{1}$ bond and is the one that has a component 
of two-spin singlet in diagonal pairs of the square 
among two possible singlet states of four spins. 
In the results in Fig.~\ref{fig6}(c) for $r=0.73$, 
the pattern of whether $\langle S_{i}^{z} S_{j}^{z} \rangle$ is 
positive or negative is common with Fig.~\ref{fig6}(b) 
and different from Fig.~\ref{fig6}(d). 
A significant difference between Fig.~\ref{fig6}(c) and Fig.~\ref{fig6}(d) 
is $\langle S_{i}^{z} S_{j}^{z} \rangle$ 
for the shortest-distant pair shown by the triangle;
its $r$-dependence is depicted in Fig.~\ref{fig6}(e) and (f). 
In Fig.~\ref{fig6}(e), 
the dependence reveals a discontinuity at $r\sim 0.675$ 
for $N_{\rm s}=36$ and 40; 
in Fig.~\ref{fig6}(f), another discontinuity appears 
for $N_{\rm s}=36$, which divides the region of $r$ into 
negative $\langle S_{i}^{z} S_{j}^{z} \rangle$ 
and positive $\langle S_{i}^{z} S_{j}^{z} \rangle$ regions. 
For $N_{\rm s}=40$ in Fig.~\ref{fig6}(f), 
$\langle S_{i}^{z} S_{j}^{z} \rangle$ changes its sign
around  $r$ similar to that for the discontinuity of $N_{\rm s}=36$, 
although $\langle S_{i}^{z} S_{j}^{z} \rangle$ for $N_{\rm s}=40$ 
is not discontinuous. 
One possible scenario for the spin state in the larger-ratio region 
is that,  
if the system forms plaquette singlets located at a local square
involving the $J_{1}$ bond, 
each plaquette singlet is 
the other singlet state which does not include 
a component of two-spin singlet in diagonal pairs of the square. 
Therefore, 
our calculations 
suggest that the state for $r=0.73$ shows a behavior 
that is different from that of the state for $r=0.68$ 
and that the behavior changes at common $r$ values 
for $N_{\rm s}=36$ and 40.

\begin{figure}[tb]
\begin{center}
\includegraphics[width=8cm]{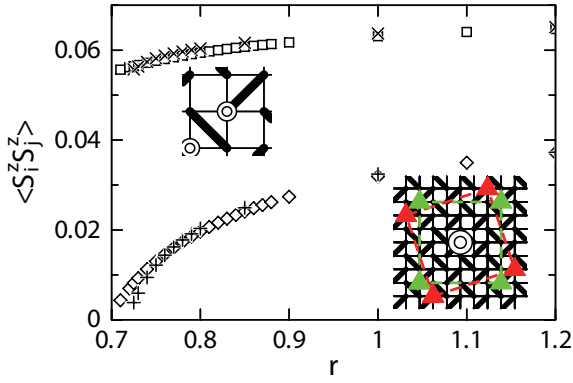}
\end{center}
\caption{(Color) 
$r$-dependence of correlation functions. 
Squares and crosses denote results of $\langle S_{i}^{z} S_{j}^{z} \rangle$ 
for next-nearest-neighbor pair illustrated in left inset. 
Diamonds and pluses denote results of $\langle S_{i}^{z} S_{j}^{z} \rangle$ 
for the longest-distance pair in the finite-size clusters. 
The right inset shows the pair from the common centered site 
by the double circle to the corner sites by green (red) closed triangles 
of the dotted-line squares for $N_{\rm s}=36$ ($N_{\rm s}=40$). 
Squares and diamonds are for $N_{\rm s}=36$; 
crosses and pluses are for $N_{\rm s}=40$.  
}
\label{fig7}
\end{figure}

\begin{figure}[tb]
\begin{center}
\includegraphics[width=8cm]{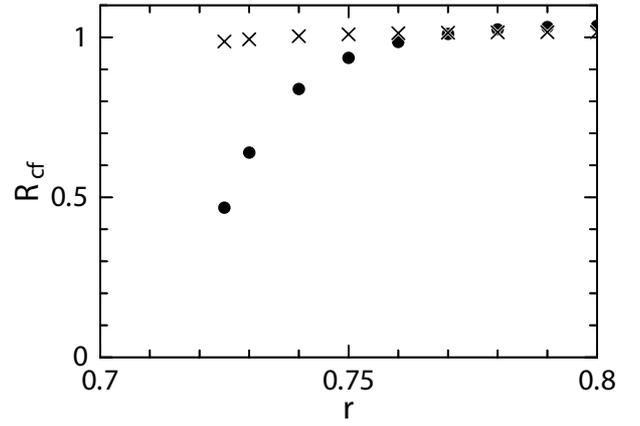}
\end{center}
\caption{
Ratios of $\langle S_{i}^{z} S_{j}^{z} \rangle$ for $N_{\rm s}=40$ 
divided by $\langle S_{i}^{z} S_{j}^{z} \rangle$ for $N_{\rm s}=36$.  
Crosses and closed circles denote 
results for the next-nearest-neighbor pair and 
for the longest-distance pair presented in Fig.~\ref{fig7}.  
}
\label{fig8}
\end{figure}

To find out 
whether or not the N$\acute{\rm e}$el-type long-range order survives, 
next, let us observe the $r$-dependence 
of correlation functions $\langle S_{i}^{z} S_{j}^{z} \rangle$ in detail; 
the results are depicted in Fig.~\ref{fig7}. 
One finds that as $r$ is decreased down to $r\sim 0.7$, 
$\langle S_{i}^{z} S_{j}^{z} \rangle$ for the next-nearest-neighbor pair 
gradually decreases, but its magnitude is not so small. 
On the other hand, $\langle S_{i}^{z} S_{j}^{z} \rangle$ 
for the pair between the longest distance decreases more rapidly; 
its magnitude becomes considerably small in the region below $r\sim 0.8$. 
To capture the difference of $\langle S_{i}^{z} S_{j}^{z} \rangle$ 
between the next-nearest-neighbor pair and the longest-distance pair, 
we examine $R_{\rm cf}$, defined as 
the ratio of $\langle S_{i}^{z} S_{j}^{z} \rangle$ for $N_{\rm s}=40$ 
divided by the corresponding $\langle S_{i}^{z} S_{j}^{z} \rangle$ 
for $N_{\rm s}=36$ presented in Fig.~\ref{fig7}; 
the results are depicted in Fig.~\ref{fig8}. 
One finds that $R_{\rm cf}$ for the longest-distance pair 
significantly decreases below $r\sim 0.76$,  
whereas the ratio for the next-nearest-neighbor pair 
is maintained at $R_{\rm cf}\sim 1$. 
This observation suggests that 
the N$\acute{\rm e}$el-type long-range order survives 
in the region above $r\sim 0.76$   
and that the order may disappear in the region below $r\sim 0.76$,  
where
the N$\acute{\rm e}$el-type short-range correlations still survive. 
As previous estimates for the edge of the N$\acute{\rm e}$el-ordered phase, 
recall $r=0.75$ in Ref.~\ref{Lou_arXiv1212_1999}, 
and $r=0.765(15)$ in Ref.~\ref{Corboz_Mila_PRB2013};   
the present result, $r\sim 0.76$ for the edge of the region 
where the N$\acute{\rm e}$el-type long-range order definitely exists, 
agrees well with those previous estimates. 

In summary, 
we have studied 
the Shastry-Sutherland model 
by the Lanczos-diagonalization method. 
The present study has presented 
diagonalization results for 36-site and 40-site 
clusters that have not been reported before. 
Our numerical results have successfully 
clarified the dependence of the spin gap 
on the ratio of interactions. 
Our calculations have successfully 
captured the edge of the dimer phase 
to be $J_{2}/J_{1}\sim 0.675$
and
the edge of the N$\acute{\rm e}$el-ordered phase   
to be $J_{2}/J_{1}\sim 0.76$. 
A noteworthy finding is a third specific ratio $J_{2}/J_{1}\sim 0.70$ 
which divides the intermediate region 
between the ratios of the two edges 
into two parts. 
We have found 
from observation of correlation functions 
that the spin state in the smaller-ratio region 
and the one in the 
larger-ratio region 
are different from each other. 
The properties of the ground states in the two intermediate regions 
should be further studied 
from different viewpoints in future. 
Such studies would greatly contribute 
to our fundamental understanding of frustrated magnetism.

\begin{acknowledgment}

We wish to thank 
Professor N.~Todoroki 
for fruitful discussions.
This work was partly supported 
by JSPS KAKENHI Grant Numbers 
16K05418, 16K05419, 16H01080 (JPhysics), and 18H04330 (JPhysics). 
Nonhybrid thread-parallel calculations
in numerical diagonalizations were based on TITPACK version 2
coded by H. Nishimori. 
In this research, we used the computational resources of the K computer 
provided by the RIKEN Advanced Institute for Computational Science 
through the HPCI System Research projects 
(Project ID: hp170018, hp170028, and hp170070). 
We used the computational resources 
of Fujitsu PRIMERGY CX600M1/CX1640M1(Oakforest-PACS) 
provided by 
Joint Center for Advanced High Performance Computing 
through the HPCI System Research project 
(Project ID: hp170207 and hp180053). 
Some of the computations were 
performed using the facilities of 
the Department of Simulation Science, 
National Institute for Fusion Science; 
Institute for Solid State Physics, The University of Tokyo;  
and Supercomputing Division, 
Information Technology Center, The University of Tokyo. 

\end{acknowledgment}

%
%
%


\begin{thebibliography}{9}
\bibitem{Shastry_Sutherland_Physica1981}
\label{Shastry_Sutherland_Physica1981}
B.~S.~Shastry and B.~Sutherland, 
Physica B \& C \textbf{108B}, 1069 (1981).
%
\bibitem{Majumdar_Ghosh_1969}
\label{Majumdar_Ghosh_1969}
C.~K.~Majumdar and D.~K.~Ghosh,
J.~Math.~Phys. \textbf{10} 1388 (1969).
\bibitem{Majumdar_1970}
\label{Majumdar_1970}
C.~K.~Majumdar, 
J.~Math.~C \textbf{3} 911 (1970).
\bibitem{Caspers_1982}
\label{Caspers_1982}
W.~J.~Caspers, Physica \textbf{115A}, 275 (1982).
\bibitem{AKLT_PRL1987}
\label{AKLT_PRL1987}
I.~Affleck, T.~Kennedy, E.~Lieb, and H.~Tasaki,
Phys.~Rev.~Lett. \textbf{59}, 799 (1987).
\bibitem{Lange_ZPhys1994}
\label{Lange_ZPhys1994}
C.~Lange, A.~Klumper, and J.~Zittartz, 
Z.~Phys.~B \textbf{96}, 267 (1994). 
\bibitem{Long_Siak_1993}
\label{Long_Siak_1993}
M.~W.~Long and S.~Siak,
J.~Phys.:~Condens.~Matter \textbf{5}, 5811 (1993). 
\bibitem{HN_VBS_PRB1996}
\label{HN_VBS_PRB1996}
H.~Nakano and M.~Takahashi, 
Phys.~Rev.~B \textbf{54}, 9000 (1996). 
\bibitem{HN_JPSJ1997}
\label{HN_JPSJ1997}
H.~Nakano and M.~Takahashi, 
J.~Phys.~Soc.~Jpn. {\bf 66}, 228 (1997).
\bibitem{HN_JPSJ1998}
\label{HN_JPSJ1998}
H.~Nakano and M.~Takahashi, 
J.~Phys.~Soc.~Jpn. {\bf 67}, 1126 (1998).
\bibitem{Tonooka_JPSJ2007}
\label{Tonooka_JPSJ2007}
S.~Tonooka, H.~Nakano, K.~Kusakabe, and N.~Suzuki,
J.~Phys.~Soc.~Jpn. {\bf 76}, 065002 (2007).
%
\bibitem{Kageyama_PRL1999}
\label{Kageyama_PRL1999}
H. Kageyama, K. Yoshimura, R. Stern, N.V. Mushnikov,
K. Onizuka, M. Kato, K. Kosuge, C. P. Slichter, T. Goto,
and Y. Ueda, Phys. Rev. Lett. 82, 3168 (1999).
\bibitem{SMiyahara_KUeda_PRL1999}
\label{SMiyahara_KUeda_PRL1999}
S.~Miyahara and K.~Ueda, Phys.~Rev.~Lett. \textbf{82}, 3701 (1999). 
%
\bibitem{AKoga_NKawakami_PRL2000}
\label{AKoga_NKawakami_PRL2000}
A.~Koga and N.~Kawakami, Phys.~Rev.~Lett. \textbf{84}, 4461 (2000). 
\bibitem{YFukumoto_JPSJ2000}
\label{YFukumoto_JPSJ2000}
Y.~Fukumoto, J.~Phys.~Soc.~Jpn. {\bf 69}, 2755 (2000).
\bibitem{Lauchli_PRB2002}
\label{Lauchli_PRB2002}
A.~L$\ddot{\rm a}$uchli, S.~Wessel, and M.~Sigrist, 
Phys.~Rev.~B \textbf{66}, 014401 (2002). 
\bibitem{Lou_arXiv1212_1999}
\label{Lou_arXiv1212_1999}
J.~Lou, T.~Suzuki, K.~Harada, and N.~Kawashima, 
arXiv:1212.1999. 
\bibitem{Corboz_Mila_PRB2013}
\label{Corboz_Mila_PRB2013}
P.~Corboz and F.~Mila, Phys.~Rev.~B \textbf{87}, 115144 (2013). 
\bibitem{Wang_Batista_PRL2018}
\label{Wang_Batista_PRL2018}
Z.~Wang and C.~D.~Batista, 
Phys.~Rev.~Lett. \textbf{120}, 247201 (2018).
%
\bibitem{HOhta_JPhysChem2015}
\label{HOhta_JPhysChem2015}
H.~Ohta, T.~Sakurai, R.~Matsui, K.~Kawasaki, Y.~Hirao, S.~Okubo, 
K.~Matsubayashi, Y.~Uwatoko, K.~Kudo, and Y.~Koike,
J.~Phys.~Chem.~B \textbf{119}, 13755 (2015). 
\bibitem{Zayed_NatPhys2017}
\label{Zayed_NatPhys2017}
M.~E.~Zayed, Ch.~R$\ddot{\rm u}$egg, J.~Larrea~J.,
A.~M.~L$\ddot{\rm a}$uchli, C.~Panagopoulos, 
S.~S.~Saxena, M.~Ellerby, D.~F.~McMorrow, Th.~Str$\ddot{\rm a}$ssle, 
S.~Klotz, G.~Hamel, R.~A.~Sadykov, V.~Pomjakushin, M.~Boehm,
M.~Jim$\acute{\rm e}$nez-Ruiz, A.~Schneidewind, 
E.~Pomjakushina, M.~Stingaciu, K.~Conder, and H.~M.~R$\o$nnow, 
Nature~Physics, \textbf{13}, 962 (2017). 
\bibitem{TSakurai_JPSJ87_ESR_HP}
\label{TSakurai_JPSJ87_ESR_HP}
T.~Sakurai, Y.~Hirao, K.~Hijii, S.~Okubo, H.~Ohta, Y.~Uwatoko, 
K.~Kudo, and Y.~Koike, 
J.~Phys.~Soc.~Jpn. \textbf{87}, 033701 (2018). 
%
\bibitem{HNakano_HaldaneGap_JPSJ2009}
\label{HNakano_HaldaneGap_JPSJ2009}
H.~Nakano, and A.~Terai, J.~Phys.~Soc.~Jpn. {\bf 78}, 014003 (2009).
%
\bibitem{HNakano_kgm_gap_JPSJ2011}
\label{HNakano_kgm_gap_JPSJ2011}
H.~Nakano and T.~Sakai, 
J.~Phys.~Soc.~Jpn. \textbf{80}, 053704 (2011).
\bibitem{HNakano_s1tri_LRO_JPSJ2013}
\label{HNakano_s1tri_LRO_JPSJ2013}
H.~Nakano, S.~Todo, and T.~Sakai, 
J.~Phys.~Soc.~Jpn. \textbf{82}, 043715 (2013).
\bibitem{HN_TSakai_kgm_1_3_JPSJ2014}
\label{HN_TSakai_kgm_1_3_JPSJ2014}
H.~Nakano and T.~Sakai, 
J.~Phys.~Soc.~Jpn. \textbf{83}, 104710 (2014).
\bibitem{HN_TSakai_kgm_S_JPSJ2015}
\label{HN_TSakai_kgm_S_JPSJ2015}
H.~Nakano and T.~Sakai, 
J.~Phys.~Soc.~Jpn. \textbf{84}, 063705 (2015).
\bibitem{HN_YHasegawa_TSakai_dist_shuriken_JPSJ2015}
\label{HN_YHasegawa_TSakai_dist_shuriken_JPSJ2015}
H.~Nakano, Y.~Hasegawa, and T.~Sakai,
J.~Phys.~Soc.~Jpn. \textbf{84}, 114703 (2015). 
\bibitem{HN_TSakai_dist_tri_JPSJ2017}
\label{HN_TSakai_dist_tri_JPSJ2017}
H.~Nakano, and T.~Sakai, 
J.~Phys.~Soc.~Jpn. \textbf{86}, 063702 (2017). 
\bibitem{HN_TSakai_tri_NN_JPSJ2017}
\label{HN_TSakai_tri_NN_JPSJ2017}
H.~Nakano, and T.~Sakai, 
J.~Phys.~Soc.~Jpn. \textbf{86}, 114705 (2017). 
\bibitem{HN_TSakai_kgm45_JPSJ2018}
\label{HN_TSakai_kgm45_JPSJ2018}
H.~Nakano, and T.~Sakai, 
J.~Phys.~Soc.~Jpn. \textbf{87}, 063706 (2018). 
\bibitem{YHasegawa_HN_TSakai_dist_shuriken_PRB2018}
\label{YHasegawa_HN_TSakai_dist_shuriken_PRB2018}
Y.~Hasegawa, H.~Nakano, and T.~Sakai,
Phys.~Rev.~B \textbf{98}, 014404 (2018). 
\bibitem{TSakai_HN_ICM018}
\label{TSakai_HN_ICM018}
T.~Sakai, and H.~Nakano, 
AIP Advances \textbf{8}, 101408 (2018). 
\bibitem{HN_TSakai_S2HaldaneGap_JPSJ2018}
\label{HN_TSakai_S2HaldaneGap_JPSJ2018}
H.~Nakano, and T.~Sakai, 
J.~Phys.~Soc.~Jpn.  \textbf{87}, 105002 (2018). 
\bibitem{You_second_deri_PRE2007}
\label{You_second_deri_PRE2007}
W.-L.~You, Y.-W. Li, and S.-J. Gu,
Phys.~Rev.~E \textbf{76}, 022101 (2007). 
\bibitem{Albuquerque_second_deri_PRB2010}
\label{Albuquerque_second_deri_PRB2010}
A. F. Albuquerque, F. Alet, C. Sire, and S. Capponi,
Phys.~Rev.~B \textbf{81},064418 (2010).
\end{thebibliography}
\end{document}